\documentclass[5p]{elsarticle}
\usepackage{lineno,amssymb,natbib}
\usepackage{float}

\makeatletter
\def\@xfootnote[#1]{%
  \protected@xdef\@thefnmark{#1}%
  \@footnotemark\@footnotetext}
\makeatother

\journal{Nuclear Instruments and Methods A}

\begin{document}

\begin{frontmatter}

\title{A Novel Transparent Charged Particle Detector for the CPET Upgrade at TITAN}

\author[TRIUMF]{D. Lascar\corref{correspondingauthor}}
\cortext[correspondingauthor]{Corresponding author. Tel: +1-604-222-1047 x6815; Fax: +1-604-222-1074}
\ead{dlascar@triumf.ca}

\author[TRIUMF,Manitoba]{B. Kootte}
\author[TRIUMF]{B.R. Barquest}
\author[TRIUMF,Manitoba]{U. Chowdhury}
\author[TRIUMF,UBC]{A.T. Gallant}
\author[TRIUMF]{M. Good}
\author[TRIUMF,MPI]{R. Klawitter}
\author[TRIUMF,UBC]{E. Leistenschneider}
\author[SFU]{C. Andreoiu}
\author[TRIUMF,UBC]{J. Dilling}
\author[TRIUMF,GRON]{J. Even}
\author[Manitoba]{G. Gwinner}
\author[TRIUMF,TAMU]{A.A. Kwiatkowski}
\author[TRIUMF,Mines]{K.G. Leach}

\address[TRIUMF]{TRIUMF, 4004 Wesbrook Mall, Vancouver, British Columbia V6T 2A3, Canada}
\address[Manitoba]{Department of Physics \& Astronomy, University of Manitoba, Winnipeg, Manitoba R3T 2N2, Canada}
\address[UBC]{Department of Physics \& Astronomy, University of British Columbia, Vancouver, British Columbia V6T 1Z1, Canada}
\address[MPI]{Max-Planck-Institut f\"{u}r Kernphysik, Heidelberg D-69117, Germany}
\address[GRON]{KVI Center for Advanced Radiation Technology, University of Groningen, Groningen, 9747 AA, Netherlands}
\address[TAMU]{Cyclotron Institute, Texas A\&M University, College Station, Texas 77843, USA}
\address[Mines]{Department of Physics, Colorado School of Mines, Golden, Colorado, 80401, USA}
\address[SFU]{Department of Chemistry, Simon Fraser University, Burnaby, British Columbia V5A 1S6, Canada}

\begin{abstract}
The detection of an electron bunch exiting a strong magnetic field can prove challenging due to the small mass of the electron. If placed too far from a solenoid's entrance, a detector outside the magnetic field will be too small to reliably intersect with the exiting electron beam because the light electrons will follow the diverging magnetic field outside the solenoid. The TITAN group at TRIUMF in Vancouver, Canada, has made use of advances in the practice and precision of photochemical machining (PCM) to create a new kind of charge collecting detector called the ``mesh detector.'' The TITAN mesh detector was used to solve the problem of trapped electron detection in the new Cooler PEnning Trap (CPET) currently under development at TITAN. This thin array of wires etched out of a copper plate is a novel, low profile, charge agnostic detector that can be made effectively transparent or opaque at the user's discretion.
\end{abstract}


\end{frontmatter}


TRIUMF's Ion Trap for Atomic and Nuclear science (TITAN) \cite{Dilling2003} is located at TRIUMF in the Isotope Separator and ACcelerator (ISAC-I) area in Vancouver, Canada. TITAN consists of 5 ion traps to further the goal of making precision mass measurements of short-lived nuclides:
\begin{enumerate}
    \item A precision hyperbolic Penning trap known as the Measurement PEnning Trap (MPET) \cite{Dilling2003}
    \item An Electron Beam Ion Trap (EBIT) \cite{Lapierre2010}
    \item A RadioFrequency Quadrupole ion cooler/buncher (RFQ) \cite{Brunner2012a}
    \item A Cooler PEnning Trap (CPET) \cite{Ke2007}
    \item A Multi-Reflection Time-of-Flight mass spectrometer (MR-ToF) \cite{Jesch2015}
\end{enumerate}
The latter two are currently being commissioned offline with the MR-ToF expected to be installed by the end of 2016 and CPET scheduled for installation roughly one year later.

Of particular interest to this work is CPET, a cylindrical Penning trap designed to cool Highly Charged Ions (HCI) delivered from the EBIT. HCI cooling in CPET occurs via the Coulomb interaction with a simultaneously trapped plasma of self-cooling electrons \cite{Ke2007}. In order to confirm the presence of trapped electrons and monitor that trapping during CPET's normal operations, a reliable detector for the electrons must be used. The CPET solenoid, with its 7 T magnetic field, strongly steers exiting electrons away from the central axis of the beamline so in order for it to be practical, the detector must be placed inside the bore of the magnet. As a result, the detector:
\begin{itemize}
    \item must remain in a fixed location in the TITAN beamline.
    \item must be effectively transparent during normal TITAN operations.
    \item should ideally be able to detect charged particles of either polarity.
\end{itemize}
This paper details the motivation, development, and commissioning of that detector.

\section{TITAN and Mass Measurements}

Ions produced at ISAC are sent in a continuous beam to the TITAN RFQ after being mass-selected by a magnetic dipole separator (mass resolution, $R\approx 2500$ \cite{Dilling2014}). In the RFQ, the ions are thermalized via a He buffer gas and bunched in a trapping volume at the end of the RFQ. Upon ejection from the RFQ, the ions---which are in a $1^+$ or $2^+$ charge state from their interaction with the He buffer gas---can be sent either directly to MPET for precision mass measurement or to the EBIT to increase their atomic charge state \cite{Leach2015}.

\begin{figure}[H]
    \begin{center}
    a)
    \includegraphics[width=0.42\textwidth]{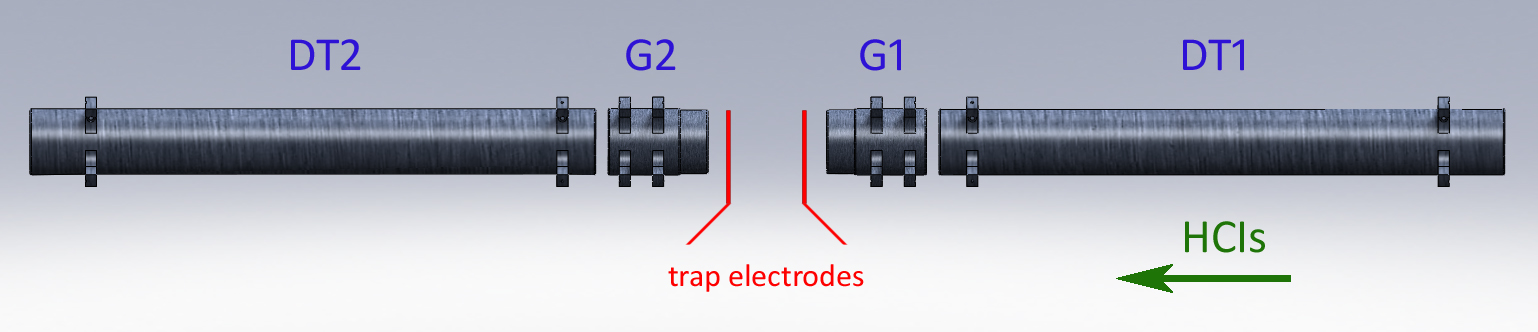}\\
    b)
    \includegraphics[width=0.42\textwidth]{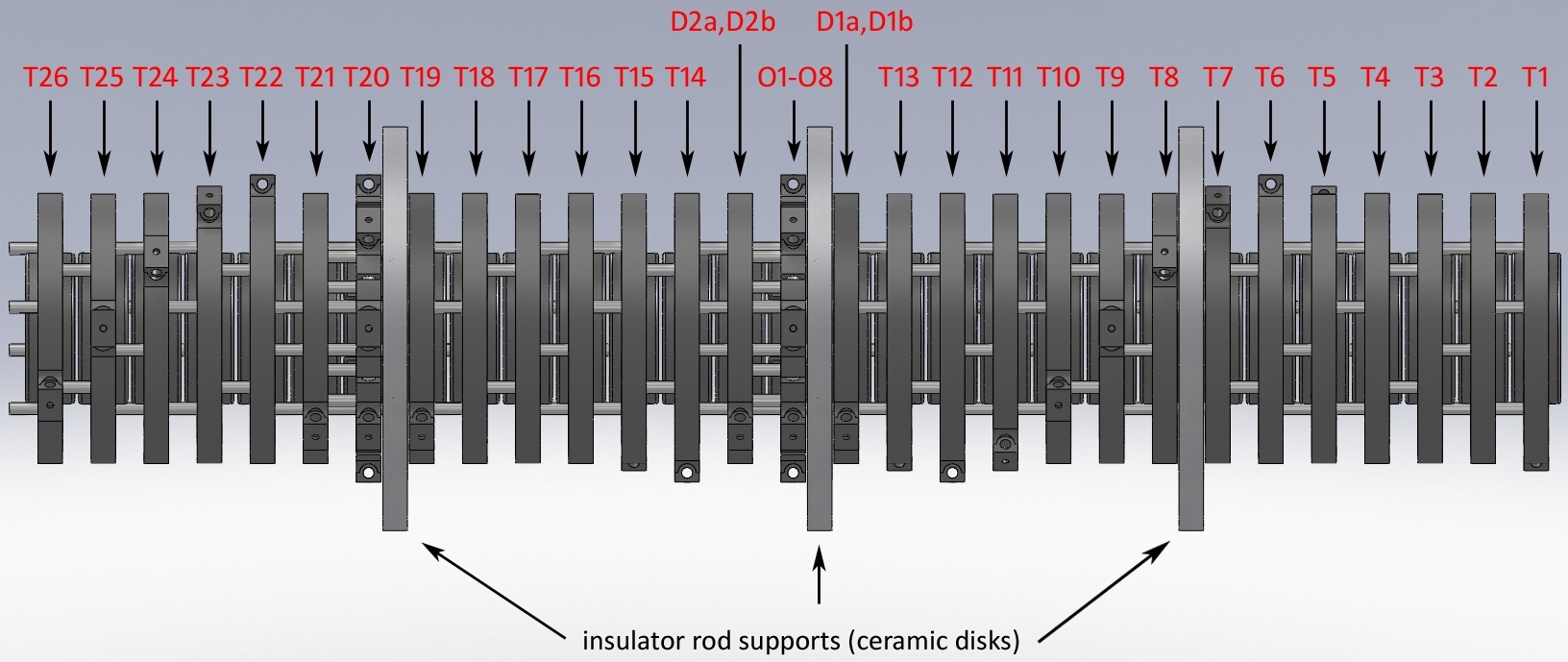}\\ 
    c)
    \includegraphics[width=0.42\textwidth]{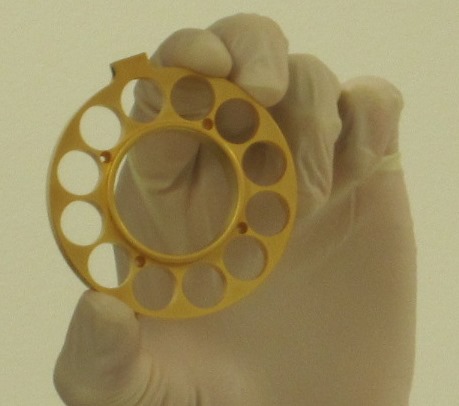}
    \caption[CPET Electrode Structure]{a \& b) A diagram of the CPET cylindrical trap electrodes \cite{Schultz2011}. On each end of the trap electrodes are so-called ``gate'' electrodes G1 and G2 whose inner diameters are identical to the central trap electrodes and are used for coarse initial trapping. All electrodes are contained in a 7 T magnetic field. a) A truncated scale image of the larger stack of electrodes surrounding the trap electrodes. The length of the entire stack, from one end of Drift Tube 1 (DT1) to the other end of Drift Tube 2 (DT2), is 120 cm. In the center of the stack (between the two red lines) are the trap electrodes. b) The trap electrodes. 29 rings of equal width whose total length (including ceramic spacers) is 40 cm. c) Closeup image of a single cylindrical electrode. The 12 smaller circles are apertures for improved pumping. For more details on the structure of CPET, please see \cite{Simon2011} (Color online).}
    \label{fig:CPET_elec}
    \end{center}
\end{figure}

MPET is TITAN's precision Penning trap which is equipped with a superconducting solenoid whose 3.7 T magnetic field confines ions radially and a harmonic electrostatic potential to confine ions axially \cite{Dilling2003,Brodeur2009,Brodeur2012}. Mass measurements in MPET are made using the Time-of-Flight Ion Cyclotron Resonance (ToF-ICR) technique which measures an ion's cyclotron frequency,

\begin{equation}
\omega_c = \frac{qeB}{m},
\end{equation}
where $B$ is the magnetic field being probed by the ion, $q$ is the absolute value of the ion's charge state, $e$ is the elemental charge, and $m$ is the ion's mass. The TITAN collaboration specializes in measuring some of the shortest-lived nuclides investigated in a Penning trap including $^{31}$Na ($t_{1/2}$ = 17 ms) \cite{Chaudhuri2013} and $^{11}$Li ($t_{1/2}$ = 8.75 ms) \cite{Smith2008}. From a technical standpoint, MPET's measurement cycle could, in its current state, be shortened to approximately 7 ms which includes a quadrupole excitation time, $t_{\mathrm{rf}}$, of 6 ms. Thus, under normal ToF-ICR experimental conditions, since the lower limit of $t_{\mathrm{rf}}$ is approximately twice the half life of the ion of interest ($t_{\mathrm{rf}} \geq 2t_{1/2}$), the lower limit on the half life of species that can be measured in TITAN is approximately 3 ms \cite{Chaudhuri2014}. This is also on the order of the limit for radioactive ion beam production through spallation from ISAC.

\subsection{Improving ToF-ICR Precision}

The fractional precision from a ToF-ICR measurement is given in \cite{Bollen2001} by,
\begin{equation}
\frac{\delta m}{m} \propto \frac{m}{q\, B\, t_{\mathrm{rf}}\, \sqrt{N_{\mathrm{ion}}}},
\label{eq:massprecision}
\end{equation}
where $N_{\mathrm{ion}}$ is the number of ions detected over the course of a measurement. The largest value for the magnetic field described in Equation \ref{eq:massprecision} is fixed by the properties of the existing superconducting solenoid. To significantly increase the field strength would require the purchase of a new magnet which would, at best given the current state of the art, yield only a factor of 2-3 improvement in precision.

As mentioned previously, the practical limit for increasing the quadrupole excitation time is $t_{\mathrm{rf}} \approx 2t_{1/2}$, so short-lived isotopes preclude longer measurement cycles that would improve precision. Many of these short-lived nuclei are weakly produced in Rare Isotope Beam (RIB) facilities, and when they are produced they are often a small fraction of a beam that contains several isobaric and molecular contaminants which reduces the overall detection sensitivity for the species of interest (see, for example, \cite{Chaudhuri2013}). As a result simply collecting more statistics to increase $\sqrt{N_{\mathrm{ion}}}$ is often impractical due to limited RIB experimental times. The charge state, $q$, however can be increased in most cases and results in an increase in the precision of a ToF-ICR measurement. The TITAN EBIT was constructed for just such a purpose and, to date, has been used to improve the precision of 30 exotic ground-state nuclei \cite{Lapierre2010,Ettenauer2011,Lapierre2012,Frekers2013,Gallant2012a,Simon2012,Macdonald2014,Chowdhury2015a,Klawitter2016} as well as an isomeric state of $^{78}$Rb$^{8+}$ \cite{Gallant2012a}.


\subsection{Motivations for CPET}

The act of charge breeding ions in the EBIT (via the process of electron impact ionization) increases the energy spread of the ion bunch sent to MPET to $\sim 10$ eV/$q$ \cite{Schupp2016}. By comparison, the cooled bunch of Singly Charged Ions (SCI) from the TITAN RFQ has a maximum energy spread of 10 eV \cite{Brunner2012a}. An increased energy spread in MPET yields a reduced signal-to-baseline measurement and thus a poorer precision, if a measurement is even possible at all. For more strongly produced ions, this can be corrected via the use of a so-called ``evaporation pulse'' which adiabatically lowers the depth of EBIT's axial trapping potential for the duration of the pulse (typically $\approx 10$ ms for the TITAN EBIT) and allows higher energy ions to drift out of the trap \cite{Lapierre2010}. In extreme cases\footnote{Such as in the measurement of $^{74}$Rb$^{8+}$ where the production of the ion of interest was greater than $10^3$ s$^{-1}$ \cite{Ettenauer2011}} this can remove up to 99.9999\% of the ion bunch in the EBIT but the remaining ions are cool enough to probe ($\sim 1$ eV/$q$). The process can be made more efficient (in terms of number of ions) if a higher threshold for evaporation is set but then a trade-off is made with the trapped bunch's energy spread.

For weakly produced species initiating a procedure that, by design, removes an inordinate number of ions of interest from the system is not desirable, so the TITAN Cooler PEnning Trap (CPET) \cite{Ke2007} was designed to cool the HCI bunch. With SCI, the cooling could be performed with a He buffer case in a gas-filled Penning trap \cite{Savard1991} or Paul trap \cite{Paul1990}, but charge states greater than $2^+$ do not survive due to charge exchange with the neutral He buffer gas. CPET is a cylindrical Penning trap and was designed to simultaneously trap a plasma of electrons and positively charged HCI by creating a set of nested electrostatic potentials in the axial direction. The electrons self-cool via the emission of synchrotron radiation in CPET's 7 T magnetic field and also cool the co-trapped HCI sympathetically via the Coulomb interaction.

The potentials are created by a stack of individually biased cylindrical electrodes (see Figure \ref{fig:CPET_elec}). The basic operating principle is to use CPET's gate electrodes (G1 and G2 in Figure \ref{fig:CPET_elec}) to trap the electrons coarsely between them and then allow the electrons to self-cool into a ``deeper'' negative potential well. Once that cooling occurs, the polarities of G1 and G2 are raised in order to trap the positively charged HCI. CPET's cooling scheme is discussed at length in \cite{Ke2007,Schultz2013,Simon2013}.

\section{Electron Detection in CPET}
\label{sec:elec_detect}

\begin{figure*}[t]
    \begin{center}
    a)
    \includegraphics[width=0.9\textwidth]{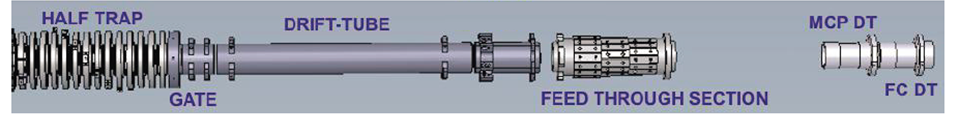}\\
    b)
    \includegraphics[width=0.9\textwidth]{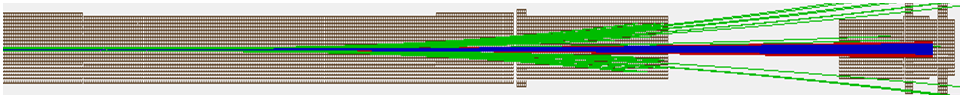}\\
    \caption[CPET Magnetic Field Decay]{CPET field and a sample beam transport simulation. a) The electrostatic components in the simulation. The electrode labeled ``GATE'' is G1 and the electrode labeled ``DRIFT-TUBE'' is DT1 as described in Fig. \ref{fig:CPET_elec}a. b) Sample SIMION\circledR\enspace simulation showing the trajectories of electrons (green), protons (red), and heavy ions with $A/q=10$ (blue) in the presence of CPET's 7 T magnetic field. The field is uniform through the center of the trap and the gate electrodes but begins to decay roughly 15 cm into DT1 (DRIFT-TUBE) such that by the time the charged particles exit DT1, the field has fallen to approximately one tenth of its maximum value. All particles began at the center of the trap in a Gaussian distribution throughout a circle of 2 mm diameter perpendicular to the beamline's central axis with an energy of 2 keV. All electrodes were set to drift tube potential (in this case, 0 V). While only 3.8(9)\% of electrons made it to the detector structures of the MicroChannel Plate (MCP) and Faraday Cup (FC) at the far right, 100\% of the protons and heavy ions reached the detector structure. The remaining 96.2\% of electrons hit trap electrodes inside the solenoid. For a) and b) while the length scales are similar, they not identical but all components in diagram \emph{a} are represented in the image from the simulation \emph{b} (Color online).}
    \label{fig:CPET_field}
    \end{center}
\end{figure*}

During CPET's initial testing phase, trapped electrons were only observed intermittently. After standard troubleshooting procedures (e.g. checking for intermittent cable connections, control signals drifting in time, failed power supplies or switches, etc.) simulations were performed with SIMION\circledR\enspace\cite{Manura2006} to check electron transmission properties. The simulations demonstrated that while successful trapping was likely occurring within the simulated experimental conditions, the detection of electrons ejected from CPET at the detector position outside the CPET solenoid would not be regular. Figure \ref{fig:CPET_field} demonstrates the effects of the diverging magnetic field on a bunch of trapped electrons 2 mm in diameter as that bunch leaves the trap.

Under the assumption that the CPET electrode stack and the magnetic field are perfectly aligned and the electron bunch is perfectly centered in the trap, the potential cross-sectional area where that spot would reach the axial position of CPET's MicroChannel Plate detectors (MCPs, see Figure \ref{fig:CPET_field}a) increases by a factor of 37. Given that the MCPs used by CPET have a diameter of 25 mm, this made detection of trapped electrons outside the magnetic field highly inefficient in the best of circumstances, and nearly impossible in the case of any trap misalignment. In most cases, the electron bunch did not interact with the MCP and was sometimes even being steered into CPET's drift tube by the magnetic field. 

\begin{figure}[H]
    \begin{center}
    a)
    \includegraphics[width=0.42\textwidth]{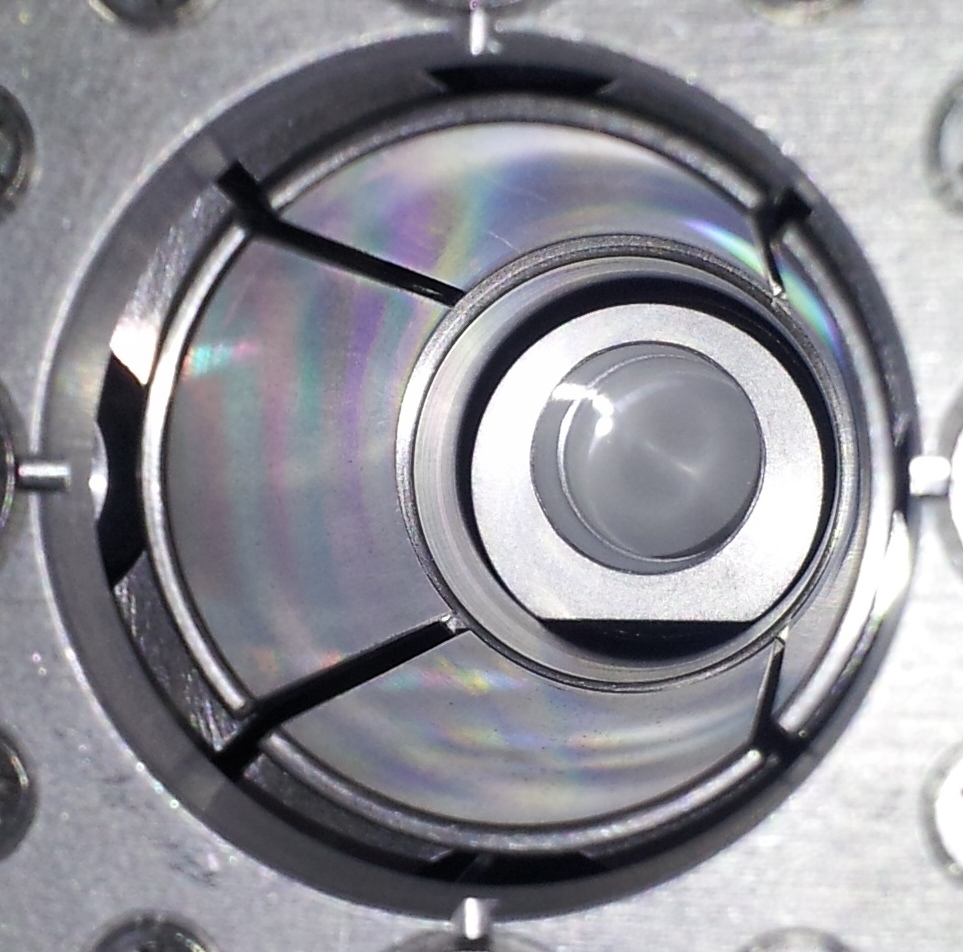}\\
    b)
    \includegraphics[width=0.42\textwidth]{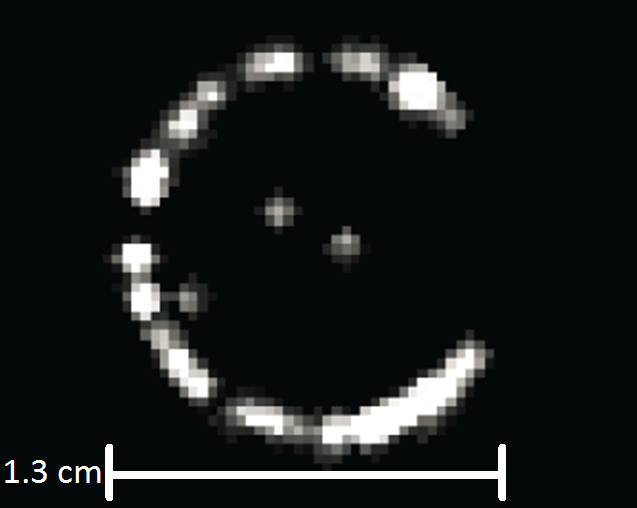}
    \caption[Diochotron plasma on the phosphor screen]{a) The CPET phosphor screen and its position in the CPET beamline. Visible in the foreground are steerer electrodes followed by the entrance to CPET's DT1 drift tube. b) CPET phosphor screen composite image (with noise reduced) of 100 electron bunches on the phosphor screen after each bunch was allowed to cool in CPET for 1.5 s. The bunches form a large ring with a diameter of approximately 1.3 cm and the size of a single bunch is illustrated in two background spots in the center of the ring. The plasma appears to be exhibiting motion under the $m=1$ diocotron plasma mode \cite{Chowdhury2015b}. (Color online)}
    \label{fig:phosphor}
    \end{center}
\end{figure}

An MCP placed 1 m away from the magnet's bore entrance will intersect with an ion bunch but will not simultaneously detect electrons. Confirmation of a trapped electron plasma is a critical diagnostic in the normal operation of CPET. Without it, any characterization of the electron trapping properties over time would be impossible and, more critically, during normal operations the experimenters would not know if any electrons were trapped at all.

The first attempt to address this issue was to place a phosphor screen (see Fig. \ref{fig:phosphor}a) in one of CPET's drift tubes (DT1, see Fig. \ref{fig:CPET_elec}a). This yielded the first results that demonstrated successful, stable electron trapping. It also yielded a stable current measurement showing that between $10^8 - 10^9$ electrons were being trapped in a single bunch in CPET, as well as the observation of an oscillating plasma mode in CPET (see Fig. \ref{fig:phosphor}b and references \cite{Chowdhury2015b,Chowdhury2015}). Despite these technical confirmations, the phosphor screen was in a position to block all ions from entering the trap, and was therefore not practical for normal CPET operation.

\section{The Mesh Detector}

\begin{figure}[h]
    \begin{center}
    a)
    \includegraphics[width=0.42\textwidth]{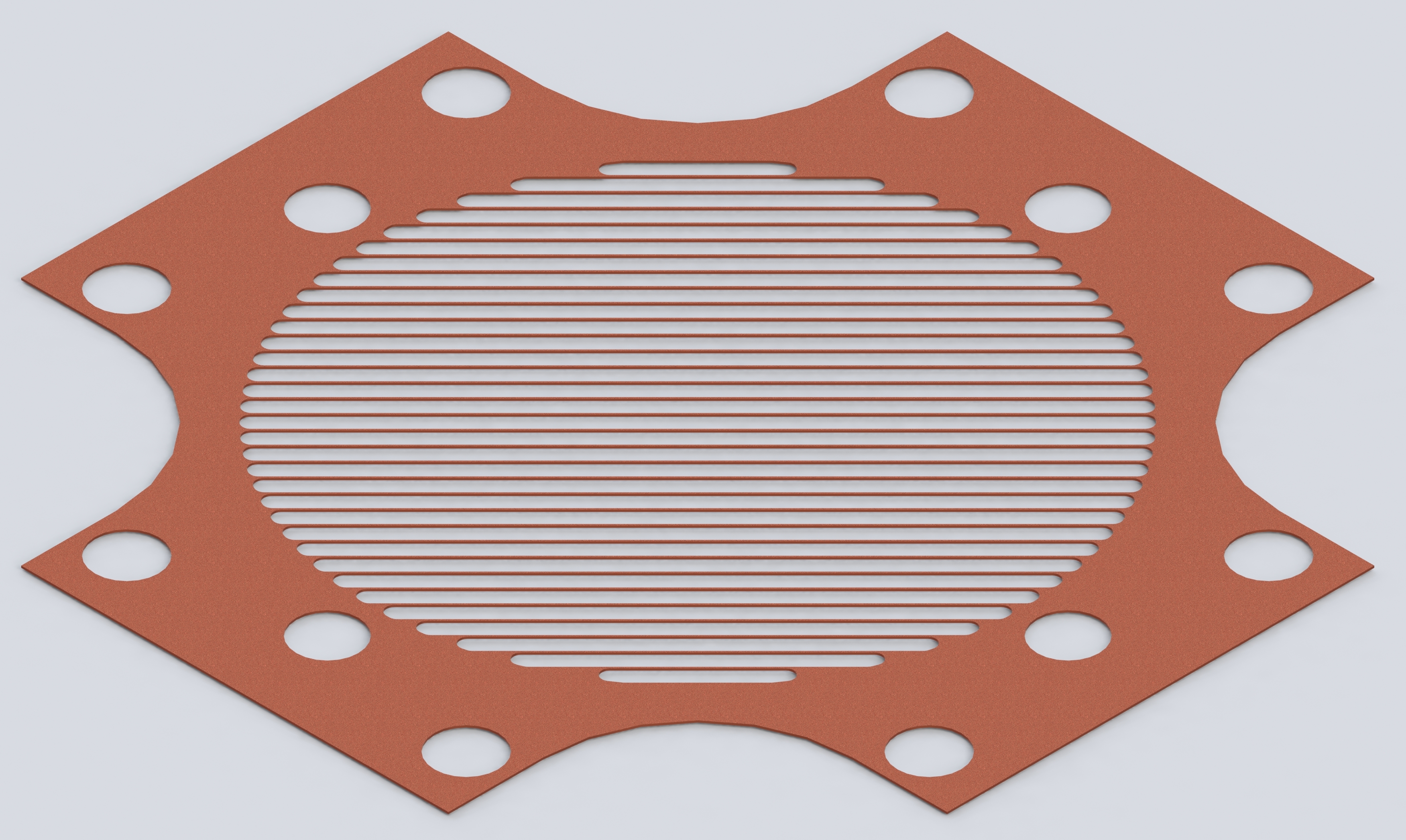}\\
    b)
    \includegraphics[width=0.315\textwidth]{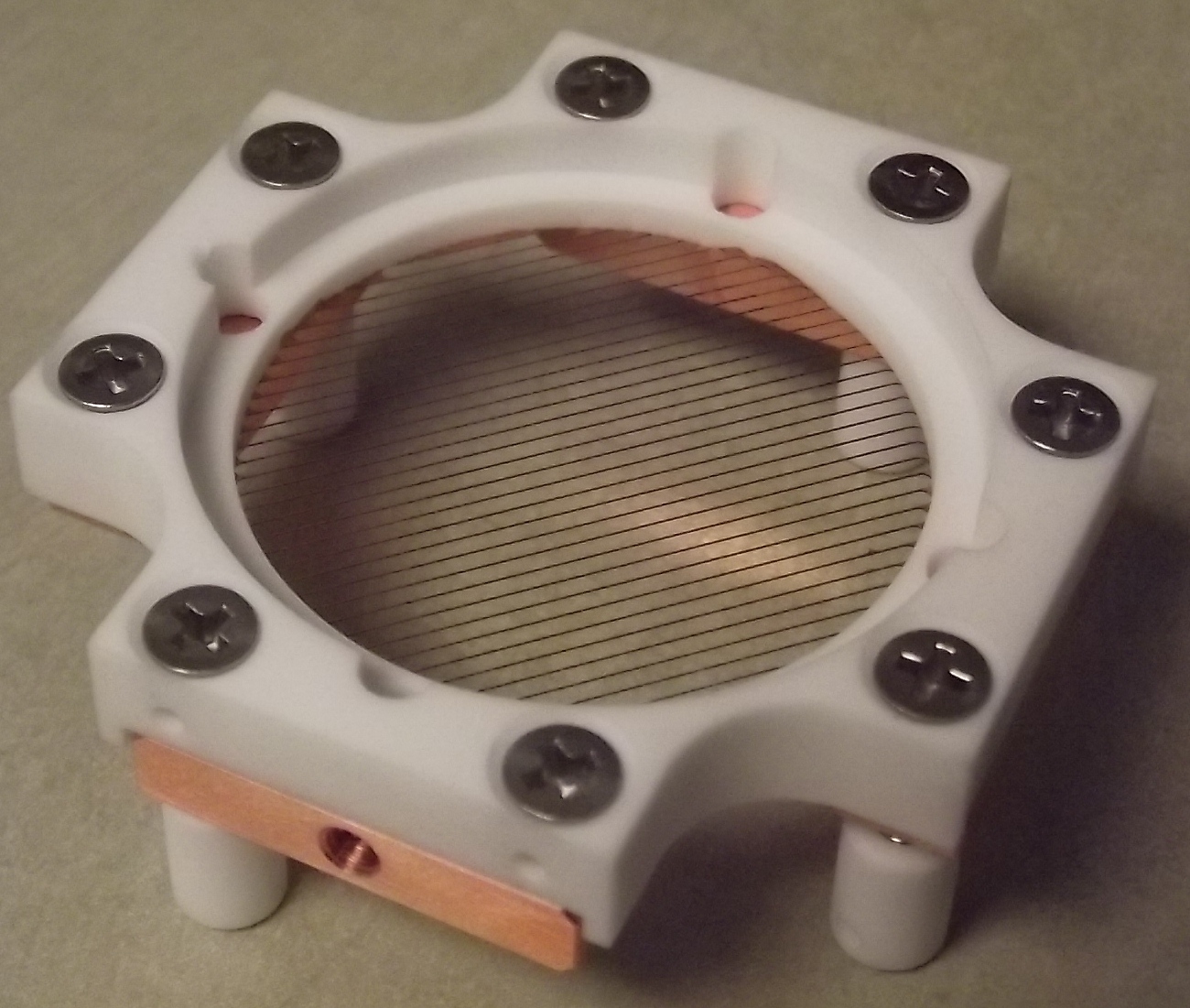}
    \includegraphics[width=0.135\textwidth]{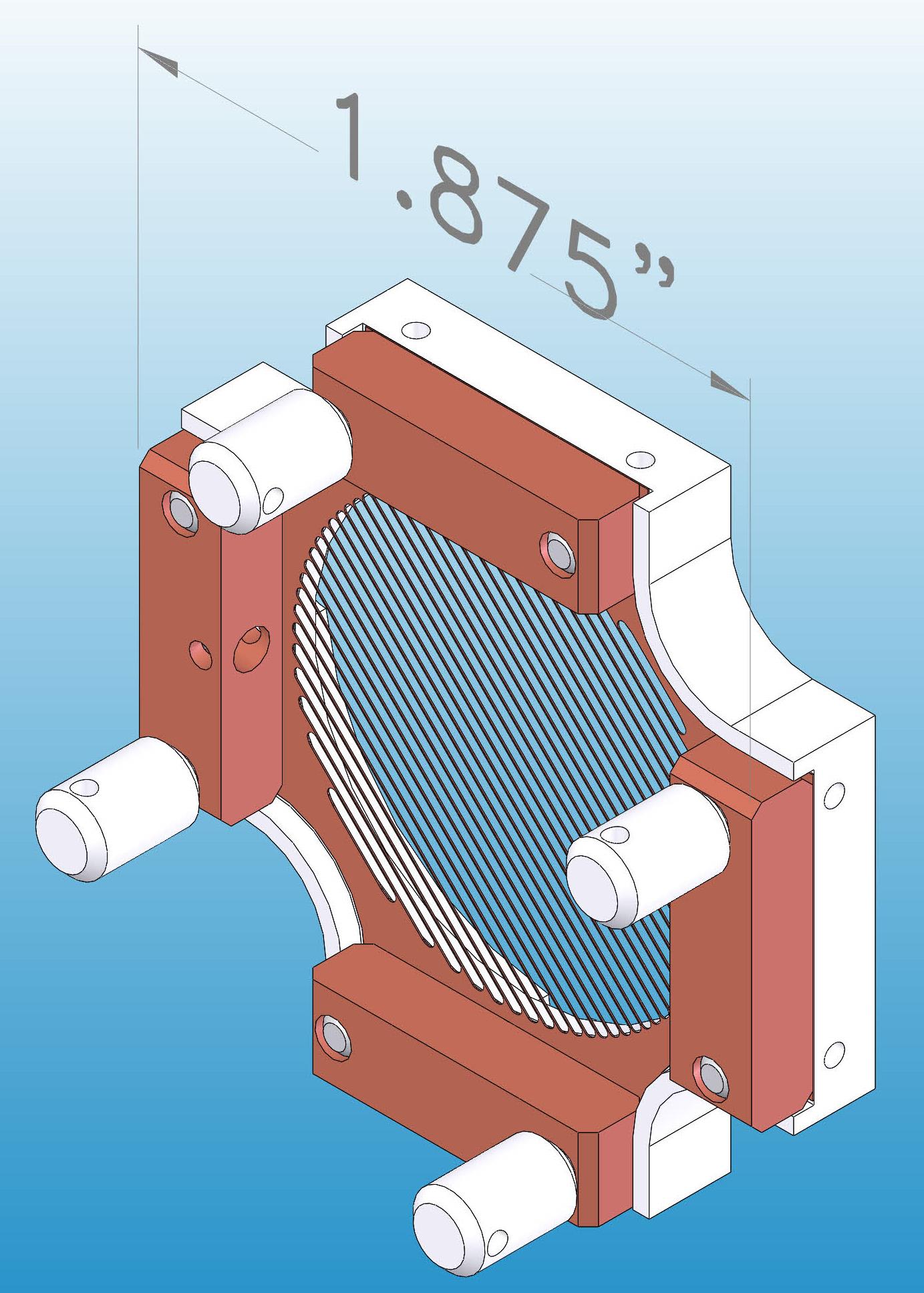}
    \caption[Mesh Detector Rendering and Photo]{a) The anode of the mesh detector. b) The assembled detector with its Copper brackets, ceramic holders, and standoffs. (Color online).}
    \label{fig:mesh}
    \end{center}
\end{figure}

A new detector was required that would not permanently block all beams entering or exiting CPET. The magnet bore of CPET is only 12.7 cm (5 inches) in diameter which makes the use of any motion feedthrough difficult if not impossible. Thus, the detector had to be a static one that could be made effectively transparent at the user's direction. The solution came by adapting TITAN's Bradbury-Nielsen Gate (BNG) design \cite{Brunner2012} for use as a new device in CPET. In a BNG, the two sets of wires are biased to
\begin{equation}
V_{\pm} = V_d \pm \Delta V_g
\end{equation}
where $V_d$ is the local drift voltage and $\Delta V_g$ is the offset added or subtracted onto a BNG plate of wires.

Just as with a BNG, if the wires are biased to a drift potential in the local region ($\Delta V_g =0$ so $V_{\pm} = V_d$), then the device is rendered effectively transparent to charged particles. Only scattering off the wires would affect a charged particle's trajectory through the detector. Whereas the BNG was made of two separate and isolated plates of wires, the new detector is made from a single plate. We have dubbed this new component the \emph{Mesh Detector}.

\subsection{Mesh Detector Design}

\begin{figure}[h]
    \begin{center}
    a)
    \includegraphics[width=0.42\textwidth]{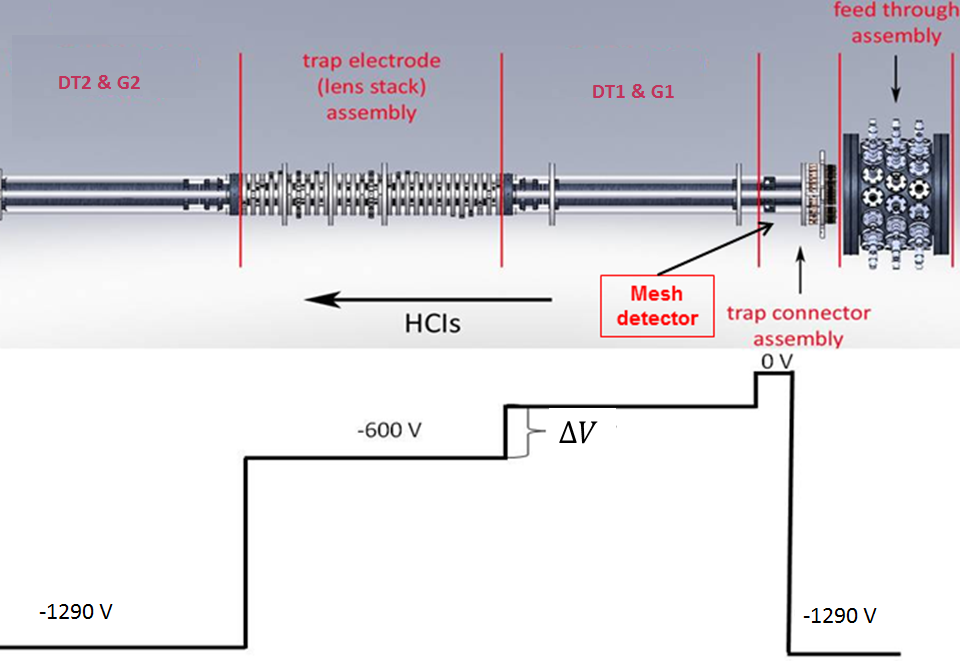}\\
    b)
    \includegraphics[width=0.42\textwidth]{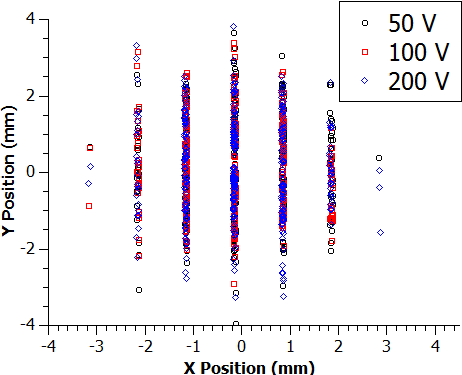}\\
    \caption[Mesh detector simulation results]{a) The biasing scheme for the mesh detector in ``opaque'' mode. When it is biased to the same potential as DT1, the detector is in ``transparent'' mode. The simulations in ``opaque'' mode were conducted at a $\Delta V$ of 50 V (-550 V on G1 and DT1) , 100 V (-500 V on G1 and DT1) and 200 V (-400 V on G1 and DT1). b) The collision position of the electrons released from a 2mm radius spherical beam spot at the center of the trap for each of the three simulated values of $\Delta V$. The distribution shows no difference in distribution as a function of $\Delta V$and appears discrete in $Y$ due to the width and pitch of the wires in the mesh detector. (Color online).}
    \label{fig:mesh_sim}
    \end{center}
\end{figure}

The detector was simulated by inserting a model of the wire plate (see Figure \ref{fig:mesh}a) into the same SIMION\circledR\enspace simulation workbench used in Figure \ref{fig:CPET_field} and discussed in Section \ref{sec:elec_detect}. The electrons began at the center of the trap, in a uniform spherical distribution with a 2 mm radius. The position was chosen to be the end of DT1 furthest from the magnet center; the same position as the phosphor screen that produced CPET's first electron detection results \cite{Chowdhury2015}. When the detector plate was biased to the same voltage as DT1 the electron collisions with an electrode surface occurred at a rate of approximately 10\% so an effective transparency of 90\% was demonstrated in the simulation. This corresponds to the detector's material coverage perpendicular to the beam's direction.

When the detector and the surrounding electrodes were biased such that the mesh detector was the ``bottom'' of an electrostatic trap, all electrons collided with the detector wires. The G2 electrode and steerer electrodes formed one of the trap walls at -1290 V, the trap electrodes were left at -600 V, DT1 was biased to create a step towards the mesh detector, and the mesh detector was set to ground. The biasing scheme is illustrated in Figure \ref{fig:mesh_sim}a. The size of the step ($\Delta V$ in Figure \ref{fig:mesh_sim}a) was varied over the course of three simulations ($\Delta V = 50, 100, 200$ V such that DT1 and G1 were biased to -550 V, -500 V, and -400 V) and Figure \ref{fig:mesh_sim}b outlines the results of those simulations by plotting the positions of the electron collisions in the radial $X$ and $Y$ directions\footnote{The $Z$ direction was the axial direction in these simulations.}. The collisions only occurred within $\pm 0.1$ mm of each other in the axial direction which corresponds to the thickness of the mesh detector plate. In the radial directions, the interactions were concentrated in strips, whose width and spacing matched those of the mesh detector wires.


It was also found that 40-50\% of the electrons collided with the detector wires within 0.1 $\mu$s. This time of flight at the energies of the electrons in the simulation was commensurate with a collision on the first pass by the wires. The remaining 50-60\% of the electrons had collision times spread out over a range of 0.2-0.4 $\mu$s which corresponds to 1-2 axial oscillations back and forth along the length of the trap until a collision occurred. As the magnitude of $\Delta V$ was increased, the fraction of electrons colliding with the detector on the first pass increased and the time spread of the remaining electrons decreased. It was initially thought that a useable signal from the detector would only include those 40-50\% so-called ``first pass'' electrons because they would be the one to provide a prompt signal. Ultimately, as will be detailed in Section \ref{sec:performance}, the distinction between ``first pass'' and the remainder of the electrons was not necessary and all electrons were counted in the signal.

The detector's main component is a 0.1 mm thick plate of Oxygen-Free High thermal Conductivity (OFHC) Copper with most of the material in a circle of 34 mm diameter removed via the process of photochemical machining (PCM) \cite{Allen2006}. What remained inside the 34 mm diameter circle is an array of parallel square wires each with a width of 0.1 mm and separated by a pitch of 1 mm. Figure \ref{fig:mesh}a shows the detector plate on its own while the remainder of Figure \ref{fig:mesh} shows the fully assembled detector with its copper clamps and ceramic holders and spacers.


Current state-of-the-art techniques claim a minimum PCM width for straight structures of 20 $\mu$m (provided that the material is at least as thick) and minimum radius of curvature of 15 $\mu$m \cite{Shimifrez2016}.

\begin{figure}[H]
    \begin{center}
    a)
    \includegraphics[width=0.42\textwidth]{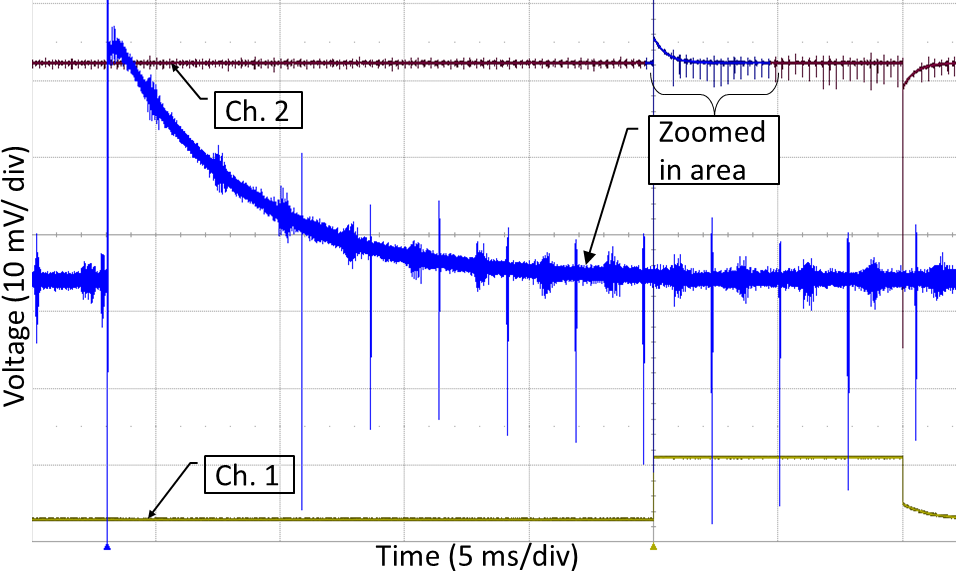}\\
    b)
    \includegraphics[width=0.42\textwidth]{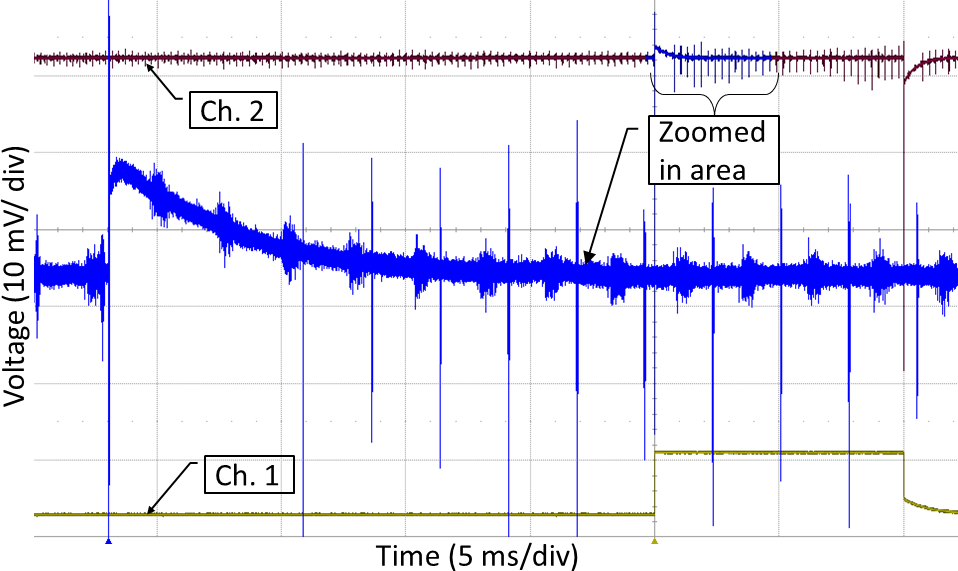}\\
    c)
    \includegraphics[width=0.42\textwidth]{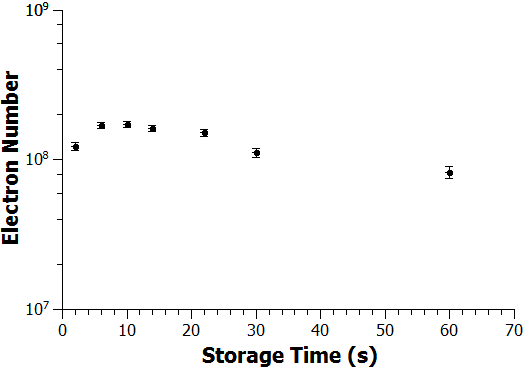}
    \caption[Electron Measurement Process]{The process of measuring electrons. a) With the Faraday Cup (FC) in place blocking electrons, the system is allowed to cycle and the integral of the noise pulse from the trap ejection in $\mu$Vs is measured. b) When the FC is removed, the electron signal suppresses the noise pulse and the difference between the two integrals yields a measure of the charge deposited on the detector. For both a) and b) the timescale is 5 ms/division. Channel 1 (yellow trace at the bottom) is the trigger for the ejection pulse, Channel 2 (red at the top) is the mesh detector signal at 10 mV/division. The blue trace is a magnification of a section of Channel 2 (bracketed in both a and b). The integrals of both traces were averaged over 100 shots. c) Measurements of electron number made with the new mesh detector as a function of time captured in CPET. Data originally appeared in \cite{Kootte2015} (Color online).}
    \label{fig:elecnum}
    \end{center}
\end{figure}

The detector anode (Figure \ref{fig:mesh}a) was placed in a custom designed MACOR\circledR\enspace holder. The anode was clamped into place via four rectangular copper clamps, also of custom design, which additionally provided the detector with biasing. The MACOR\circledR\enspace holder was designed to slide snugly over CPET's DT1 (see Figure \ref{fig:CPET_elec}a) in one direction, and in order to maintain electrical isolation, commercially available MACOR\circledR\enspace standoffs were attached. The complete detector assembly can be seen in Figure \ref{fig:mesh}b.


In order to accommodate the mesh detector and its attendant support structure, approximately 1 inch of DT1 was removed so the detector could occupy the space. Also, CPET's T26 electrode was connected to the G2 electrode (see Figure \ref{fig:CPET_elec}a and b) inside the vacuum so that an electrical feedthrough in the CPET feedthrough section (see Figure \ref{fig:CPET_elec}a and b) could be made available for the mesh detector.


\subsection{Mesh detector performance}
\label{sec:performance}

After installation, the performance of the detector was checked. Electrons were loaded into CPET with the electrostatic trap walls formed by the two gate electrodes (G1 and G2). With the trap in the ``closed position'' the gate electrodes were each biased to -1290 V. The central trap electrodes were all biased to -630 V so the entire trap volume was used to hold electrons. When the trap was opened to eject the electrons towards the detector, the G1 electrode was pulsed to +100 V and the electrons were allowed to drift towards the detector.

The charge deposited on the detector was read by integrating the signal on a 500 MHz oscilloscope. Noise on the detector was a problem because the 1400 V extraction pulse on G1 generated an image charge on the detector which overlapped with the calculated time of the electrons' arrival. In order to solve this problem, the oscilloscope channel monitoring the detector was set to a 1 M$\Omega$ coupling to ground and the CPET system was cycled in two modes:

\begin{enumerate}
    \item Electrons allowed to pass into the trap.
    \item Electrons blocked via the insertion of a Faraday cup into the beamline directly in front of the electron source.
\end{enumerate}

In both modes, all capture and ejection pulses were applied as normal but in mode 2, a Faraday cup was placed in the path of the electrons to block their entrance to the trap. Following this, it was possible to determine that a signal was present above the noise background. With no electrons and with a 1 M$\Omega$ coupling on the oscilloscope channel, a roughly 3 mV pulse decaying with a time constant of approximately 300 $\mu$s was observed. With electrons, that pulse dropped in height to $\sim 1$ mV at the same decay constant. The signals were averaged over 100 shots and the uncertainty in the integral was estimated via observation of the variation of the integral once 100 shots had been reached.

To measure the charge deposited, the integrals of the signals were averaged in each of the 2 modes and the two integrated oscilloscope traces were subtracted from each other. The differences in the integrals, $\Delta A$ in units of V$\cdot$s yielded a total charge deposited, $q$, via:
\begin{equation}
q = \frac{\Delta A}{eR},
\label{eq:charge}
\end{equation}
where $R$ is the circuit's resistance. In this case, $R$ is dominated by the 1 M$\Omega$ coupling to ground. In general, $A \sim 100$ $\mu$Vs and $\Delta A \sim 10$ $\mu$Vs. The number of captured electrons was calculated from Equation \ref{eq:charge} and plotted in Figure \ref{fig:elecnum}c) as a function of their time captured in CPET. The uncertainty in the electron number measurement ($\approx 7.5 \times 10^6$ $q$) is plotted as well. What was observed was an overall electron number on the order of $10^8$. Within that order of magnitude, we observed a small rise of less than a factor of 2 between bunches stored in CPET for 2 and 6 seconds followed by a slow decay at longer times. We believe that the initial rise in detection is due to the damping of the collective diochotron plasma motion that brings the electron plasma towards the center of the trap \cite{Kootte2015} that we have observed before \cite{Chowdhury2015}. The slow decay at longer storage times is due to collisions between electrons and residual neutral gas in the CPET volume. These results match electron number measurements mentioned at the end of Section \ref{sec:elec_detect} where the phosphor screen (whose active area overlapped with 100\% of the possible electron paths) was used as a charge collecting anode.

With the properly calculated electron number and the cooling curves for HCI simulated in Figure 2 of Ke at al. \cite{Ke2007} the cooling time can be estimated. For example, when Ke et al. assumed a nested trapping volume within six of the 29 CPET trap electrodes and an upper limit of trapped $10^3$ HCI, the author demonstrated that cooling times of less than 100 ms can be achieved.

\section{Outlook and Conclusion}

The TITAN mesh detector has been developed and has demonstrated proper electron detection inside CPET's strong magnetic field. It allows experimenters to take advantage of advances in precision and boutique-level customization provided by the PCM process. The mesh detector's simplicity and low material cost makes it a cheap, robust option for many experiments. The detector's design is quite rugged and can be adapted to almost any experimental location with little or no modification to the experiment because of its small profile. Wire configuration, spacing, size and thickness can all be specified by the user. Material choice is also an option for potentially corrosive environments. The final detector is charge-agnostic, and with a careful choice of biasing, can be made effectively transparent to passing charged particles during normal experimental operations when beam diagnostics are not needed.

At CPET, the mesh detector has been demonstrated to determine reliably the number of trapped electrons as well as to switch rapidly between measurement and transparent modes. Consequently, the setup described herein will be improved upon for its final and permanent implementation. Nevertheless, the initial setup has allowed the CPET project to progress towards the goal of simultaneous capture of electrons and HCI followed by HCI cooling.

\section{Acknowledgements}
TRIUMF receives federal funding via a contribution agreement with the National Research Council of Canada (NRC). This work was partially supported by the Natural Sciences and Engineering Research Council of Canada (NSERC), the Canada Foundation for Innovation (CFI) and the Deutsche Forschungsgemeinschaft (DFG) under Grant FR 601/3-1. BK and UC acknowledge the support from the University of Manitoba Faculty of Science Scholarships. DL wishes to thank C. Jannace and F. Friend for editing support.

\section*{References}

\bibliography{library}

\end{document}